\UseRawInputEncoding
\documentclass[12pt]{spieman}  
\usepackage{amsmath,amsfonts,amssymb}
\usepackage{graphicx}
\usepackage{cleveref}
\usepackage{setspace}
\usepackage[modulo]{lineno}
\usepackage{lipsum}
\usepackage{tocloft}
\usepackage{color}
\usepackage{soul}

\title{Analyzing the Atmospheric Dispersion Correction of the Gemini Planet Imager: residual dispersion above design requirements}
\author[a,b, *]{Malachi Noel}
\author[a]{Jason J. Wang}
\author[c, d, e]{Bruce Macintosh}
\author[f]{Katie Crotts}
\author[f, g]{Christian Marois}
\author[h]{Eric L. Nielsen}
\author[i]{Robert J. De Rosa}
\author[a]{Katie Scalzo}
\author[j]{Kent Wallace}

\affil[a]{Center for Interdisciplinary Exploration and Research in Astrophysics (CIERA) and Department of Physics and Astronomy,
Northwestern University, Evanston, IL 60208, USA}
\affil[b]{New Trier High School, 385 Winnetka Ave, Winnetka, IL 60093, USA}
\affil[c]{Kavli Institute for Particle Astrophysics and Cosmology, Stanford University, Stanford, CA 94305, USA}
\affil[d]{University of California Observatories, 1156 High Street, Santa Cruz, CA 95064, USA}
\affil[e]{ Department of Astronomy and Astrophysics, University of California, Santa Cruz, Santa Cruz, CA 95064, USA}
\affil[f]{University of Victoria, 3800 Finnerty Road, Victoria, BC, V8P 5C2, Canada}
\affil[g]{National Research Council of Canada Herzberg, 5071 West Saanich Rd, Victoria, BC, V9E 2E7, Canada}
\affil[h]{Astronomy Department, New Mexico State University, 1320 Frenger Mall, Las Cruces, NM 88003, USA}
\affil[i]{European Southern Observatory, Alonso de C\'{o}rdova
3107, Vitacura, Santiago, Chile}
\affil[j]{Jet Propulsion Laboratory, California Institute of Technology, 4800 Oak Grove Dr.,Pasadena, CA 91109, USA}

\cftpagenumbersoff{figure}
\cftpagenumbersoff{table} 
\begin{document} 
\maketitle

\begin{abstract}
The Atmospheric Dispersion Corrector (ADC) of the Gemini Planet Imager (GPI) corrects the chromatic dispersion caused by differential atmospheric refraction (DAR), making it an important optic for exoplanet observation. Despite requiring less than 5 mas of residual DAR to avoid potentially affecting the coronagraph, the GPI ADC averages $\sim7$ and $\sim11$ mas of residual DAR in $H$ and $J$ band respectively. We analyzed GPI data in those bands to find explanations for the underperformance. We found the model GPI uses to predict DAR underestimates humidity's impact on incident DAR, causing on average a 0.54 mas increase in $H$ band residual DAR. Additionally, the GPI ADC consistently undercorrects in $H$ band by about 7 mas, causing almost all the $H$ band residual DAR. $J$ band does not have such an offset. Perpendicular dispersion induced by the GPI ADC, potentially from a misalignment in the prisms' relative orientation, causes 86\% of the residual DAR in $J$ band. Correcting these issues could reduce residual DAR, thereby improving exoplanet detection. We also made a new approximation for the index of refraction of air from 0.7 microns to 1.36 microns that more accurately accounts for the effects of humidity.
\end{abstract}


{\noindent \footnotesize\textbf{*}Corresponding author: Malachi Noel,  \linkable{malachidnoel@gmail.com}. }

\begin{spacing}{1}   

\section{Introduction}

\label{sect:intro}  
Light disperses when entering our atmosphere so that shorter wavelengths which generally refract more appear closer to the zenith. Different wavelengths of light from the same object appear in different places as a result. Consequently, an object's location changes as it's observed across different wavelengths of light. Differential Atmospheric Refraction (DAR) refers to this effect of wavelength dependent refraction. The refractive index of air at a given wavelength, and thus incident DAR, depends on temperature, pressure, and humidity. Additionally, zenith distance affects the degree of dispersion, with more dispersion at greater zenith distances and no dispersion at the zenith. {\cite{Roe2002}} DAR is particularly observable when using an Integral Field Spectrograph (IFS) behind adaptive optics. In IFS data, objects appear to drift across different wavelengths due to DAR. To combat dispersion, an Atmospheric Dispersion Corrector (ADC) can be used to reverse the differential refraction. 

\subsection{The Gemini Planet Imager (GPI) ADC}
\label{subsec:GPI}

The GPI and its ADC had their first light in November 2013 {\cite{GPIFirstlight}}. The GPI ADC essentially acts as a prism to reconverge dispersed light. It is a double prism ADC that has three mechanical degrees of freedom: by increasing the prism separation, the ADC is able to compensate for larger dispersion; the orientation of the double prism assembly is rotated to align with the elevation axis of the telescope; the relative orientation of the two prisms remains fixed during the entire time although each prism is able to rotate individually. It considers temperature, pressure, and humidity to calculate the refractive index of air. Using the index of refraction of the air and the zenith distance, the GPI ADC calculates the expected dispersion and correction needed. The GPI ADC is designed to operate at zenith distances less than 50$^\circ$, hence why its performance drops off sharply thereafter.{\cite{GPIADC}}

The GPI currently uses a refractive index of air model very similar to Roe's model{\cite{Roe2002}}, which calculates the refractive index of air using pressure, temperature, humidity, and wavelength. The Mathar model{\cite{Mathar}} is another refractive index of air model that considers pressure, temperature, humidity, and wavelength. At a temperature of 13$^\circ$C, pressure of 548 mm Hg, relative humidity of 25\%, and zenith distance of 26$^\circ$, the Mathar model estimates the incident DAR in the {$H$} band to be 15.39 mas. Under these conditions, the Roe model's estimate is slightly lower, at 14.92 mas. However, if the relative humidity is adjusted to 0\%, the Mathar model's prediction changes to 14.86 mas, and the Roe model forecasts a nearly identical incident DAR of 14.90 mas. As the Mathar first-principles model is difficult to run, several approximations exist at discrete wavelength ranges, including one that operates from 1.3 $\mu$m to 2.5 $\mu$m{\cite{Mathar}}. We also created a new approximation that is valid from 0.7 $\mu$m to 1.36 $\mu$m which can be found in Appx. \ref{appx:approx}. Previously, no Mathar approximation existed for the $Y$ and $J$ bands. References to the Mathar model in this paper refer to whichever Mathar approximation is valid in the context. Many other models also exist for the refractive index of air. We used the AstroAtmosphere python module{\cite{astroatmo}} to access models other than the Roe and Mathar models.

\begin{figure}
\begin{center}
\begin{tabular}{c}
\includegraphics[height=6.5cm]{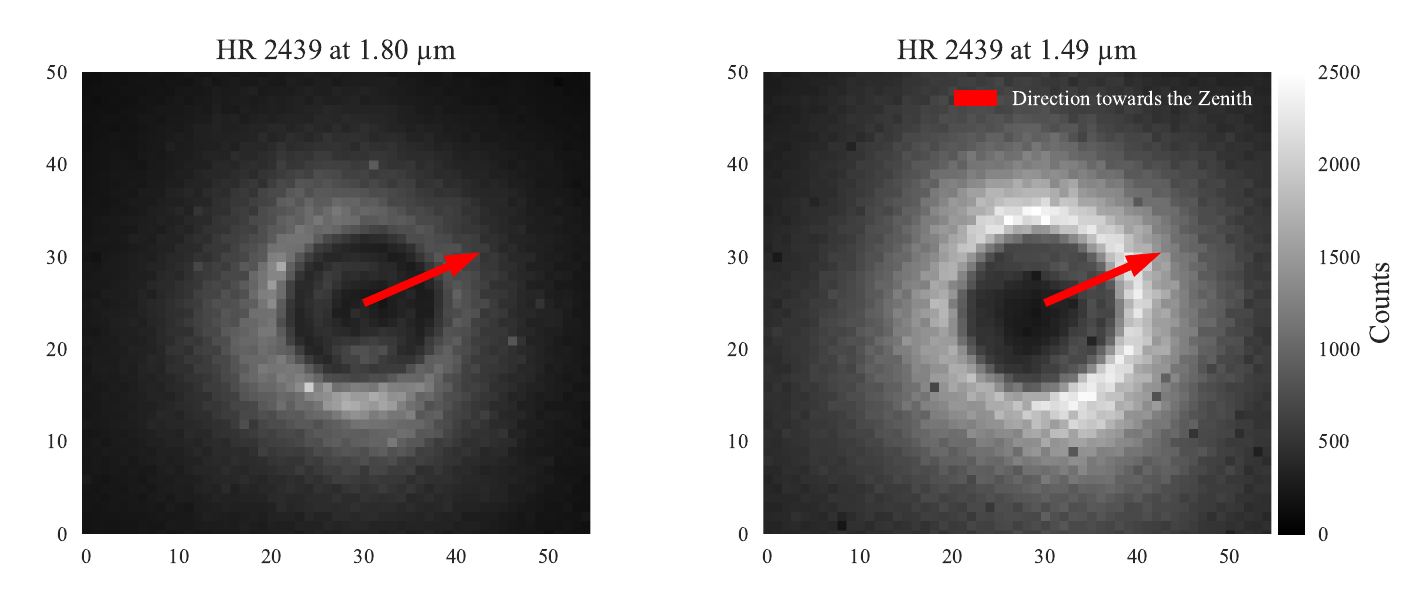}  
\\
(a) \hspace{7.2cm} (b)
\end{tabular}
\end{center}
\caption 
{ \label{fig:dispersiondemo}
Visual demonstration of the effects of residual DAR from (a) 1.80 $\mu$m to (b) 1.49 $\mu$m in HR 2439 GPI IFS $H$ band data. The star drifts towards the zenith with decreasing wavelength, causing it to move out from behind the coronagraph. At 1.49 $\mu$m, the star is no longer positioned well behind the coronagraph, allowing light from the star to drown out potential light from exoplanets.} 
\end{figure}

At design, the GPI ADC aimed to reduce dispersion to less than 5 mas. An accuracy of 5 mas is ideal for coronagraphy, and an accuracy of 1 mas is ideal for astrometry.{\cite{GPIADC}} Figure \ref{fig:dispersiondemo} shows a high zenith distance cube of HR 2439 with an anomalously high 23.0 mas of residual dispersion to demonstrate how residual DAR affects coronagraphy. The figure shows an extreme case for demonstrational purposes at zenith distances beyond those at which the ADC operates. The high residual DAR in this spectral datacube significantly impacts its quality as the star does not remain well positioned behind the coronagraph. Reducing dispersion on the GPI would improve the detection of exoplanets at small angular separations that are close to the coronagraphic mask. In this paper, we will analyze the performance of the GPI ADC and explore its areas for improvement to further our ability to observe exoplanets using GPI. 

\section{Methodologies and Data}
\subsection{Observations and Spectral Datacube Processing}
To characterize the GPI ADC, we used all on-sky integral field spectroscopy taken in $J$ and $H$ band as part of the GPI Exoplanet Survey (GPIES; GS-2014B-Q-500, GS-2014B-Q-501, GS-2014B-Q-503, GS-2015A-Q-500, GS-2015A-Q-501, GS-2015B-Q-500, GS-2015B-Q-501, GS-2016A-Q-500, GS-2016A-Q-501, GS-2016B-Q-500, GS-2016B-Q-501, GS-2017A-Q-500, GS-2017A-Q-501, GS-2017B-Q-500, GS-2017B-Q-501, GS-2019A-Q-500) {\cite{Nielsen2019}}. In total we collected 25,566 exposures in $H$ band and 1,129 exposures in $J$ band. Note that GPIES primarily observed in $H$ band, which explains the order of magnitude more frames in $H$ band. We did not consider $Y$ and $K$ band due to the limited amount of data at those wavelengths.

The raw data was processed with the GPIES Data Cruncher framework and the GPI Data Reduction Pipeline following its standard processing steps for integral field spectroscopy {\cite{Perrin2014, Perrin2016, Wang2018}}. We briefly summarize the steps here. The raw data is dark subtracted and bad-pixel corrected. Then, a wavelength calibration is used to extract all the microspectra on each detector frame, resulting in a 3-D spectral data cube containing an image of a star behind the coronagraph at 37 separate spectral channels. The images are corrected for optical distortion {\cite{Konopacky2014}} before astrometry is performed. To measure the position of the star behind the coronagraph at each wavelength, we measure the position of the four ``satellite spots" that are generated by a diffractive pupil grid and centered on the position of the occulted star {\cite{Satspots,Satspots2,Satspots3}}. The positions of the four satellite spots in each spectral channel are recorded into the header of each spectral datacube. 

\subsection{Data Taken From Datacube Headers}
\label{subsec:headers}
Almost all of the data analyzed in this paper were taken from the headers of fits files. We used the header keyword  ``PRESSURE" to get the atmospheric pressure in mm Hg, ``TAMBIENT" to get the temperature in $^\circ$C, ``HUMIDITY" to get the relative humidity, and ``ELEVATIO" to get the elevation in degrees. Additionally, GPI creates satellite spots, artifacts created by a diffraction grid that indicate the exact location of a star. The average of the coordinates of the four satellite spots in the field of view is the coordinates of the star.{\cite{Satspots,Satspots2,Satspots3}} The coordinates for the $j$th satellite spot in the $i$th frame can be found using the header keyword ``SATS$i$\_$j$" in the extension header. The first frame in each spectral datacube and the first satellite spot in each frame are referred to as the 0th frame and 0th satellite spot respectively. The CD matrix is available in the extension header using the keywords ``CD1\_1", ``CD1\_2", ``CD2\_1", and ``CD2\_2." We used the CD matrix which is derived from the telescope pointing information to derive the elevation axis on the GPI detector, which we will use in Sec. \ref{subsec:analyzecorrection}.
\subsection{Measuring Differential Atmospheric Refraction}
\label{subsec:measuredrift}

\begin{figure}
\begin{center}
\begin{tabular}{c}
\includegraphics[height= 9cm]{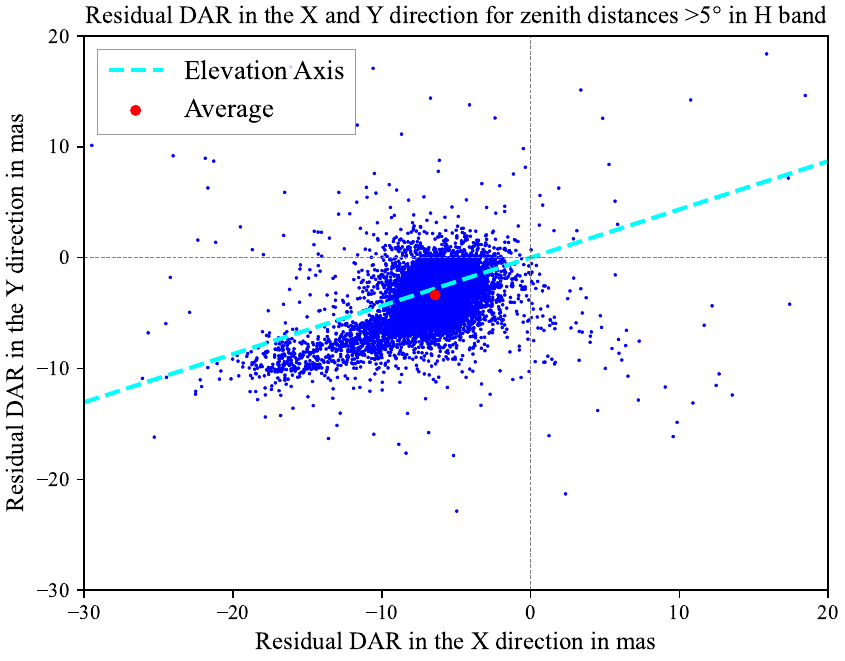}
\end{tabular}
\end{center}
\caption 
{ \label{fig:dispersioncoord}
$H$ band residual DAR in the X and Y direction. The dispersion in each direction is calculated as described in Sec. \ref{subsec:measuredrift}. The red dot shows the average residual DAR in the data. These data points show DAR with increasing wavelength, so the incident DAR goes away from the zenith. The zero point is where the best-fit lines are flat for the x-coordinate vs wavelength and y-coordinate vs wavelength graphs. } 
\end{figure}

The residual DAR measured in each spectral datacube is the DAR present after the ADC applies its correction. We used the coordinates of the satellite spots to determine the position of the star. After we determined the center of each frame in a spectral datacube, we then fit a line for the location of the X coordinate as a function of wavelength and took the difference between the first and last point on the best-fit line to be the average residual X-direction DAR. We then repeated the same operation for the Y coordinates to get the average residual DAR for each spectral datacube. Across all datacubes, there was an average root-mean-square deviation of 0.83 mas and 0.81 mas in the X-direction and Y-direction respectively. Figure \ref{fig:dispersioncoord} shows a comparison of residual X-direction DAR and residual Y-direction DAR. Using a best-fit line to get the residual DAR instead of finding the difference in the center from the first and last frame reduces the effect of instrumental noise. We then converted the residual DAR from pixels into mas using the GPI plate scale of 14.161 mas/px {\cite{GPIScale}}. Since we calculated residual DAR by subtracting the smaller wavelength data (1.49 $\mu$m in $H$ band and 1.1 $\mu$m in $J$ band) from the larger wavelength data (1.80 $\mu$m in $H$ band and 1.35 $\mu$m in $J$ band), the residual DAR direction is away from the zenith in the data we analyze in this paper as we are measuring how the star moves as wavelength increases. Subtracting the larger wavelength data from the smaller wavelength data would result in the residual DAR being in the direction of the zenith as it would measure how the star moves as wavelength decreases. 
\subsection{Pruning the GPI IFS Data}
\label{subsec:prune}
We accessed 25,566 spectral datacubes in $H$ band taken from November 2014 to August 2019 at the Gemini South Telescope. Almost all of the data were taken at temperatures from 3 to 19$^\circ$C, pressures from 543 to 551 mm Hg, relative humidities from 7 to 74\%, and zenith distances from 0 to 60$^\circ$. We removed all spectral datacubes whose headers indicated an ADC angle of 0 as that means the ADC was likely not deployed properly. Spectral datacubes without a pressure reading were also removed. One spectral datacube had a temperature close to 0$^\circ$C and was removed because its temperature reading differed significantly from the rest of the data. Information on ADC angle, temperature, and pressure are all available in the headers as described in Sec. \ref{subsec:headers}. After pruning bad data, we were left with 24,314 spectral datacubes in $H$ band. 

We accessed 1,129 spectral datacubes in $J$ band taken from December 2014 to November 2018 and were left with 1,118 spectral datacubes after pruning them the same way we pruned $H$ band data. 
\begin{figure}
\begin{center}
\begin{tabular}{c}
\includegraphics[height=9cm]{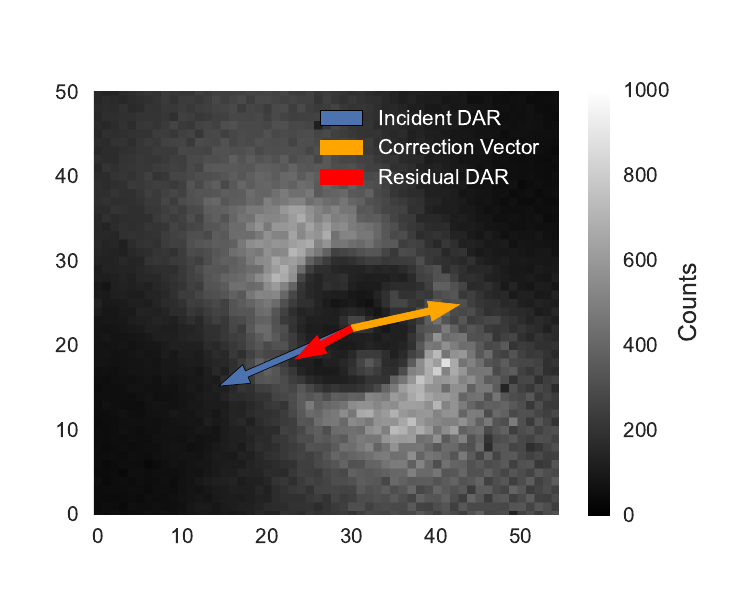}  
\end{tabular}
\end{center}
\caption
{ \label{fig:vectors}Visualization of the modeling of residual DAR, incident DAR, and ADC correction on a 51 Eridani spectral datacube. The vectors show the DAR with increasing wavelength, so the incident DAR goes away from the zenith. Vector magnitudes are magnified by a factor of 10 for demonstration. The correction vector's angle offset is amplified by a factor of 5. } 
\end{figure} 
\subsection{Model and Correction Vectors}
\label{subsec:getcorrection}

To analyze the GPI ADC, we need a measure of the corrections it applies. We can model the incident DAR, the correction applied by the ADC, and the residual DAR as 2-dimensional vectors. Figure \ref{fig:vectors} visualizes these three vectors. The residual DAR is essentially the incident DAR vector plus the correction vector. Since the correction vector is in the opposite direction of the atmospheric dispersion, they nearly cancel each other out. As the artificial source is downstream from the ADC in GPI, we cannot measure the true correction vector using internal calibration sources, so we attempted to find the inferred correction vector. Switching terms around, we can find the inferred correction vector by subtracting the atmospheric dispersion from the residual DAR. To do this, we need to model the atmospheric dispersion. Models for the refractive index of air allow us to calculate the expected dispersion for each spectral datacube using atmospheric variables. We used a polynomial approximation of the Mathar model valid in $H$ band{\cite{Mathar}} to calculate the refractive index of the air outside of the dome at the time of observations for $H$ band data. According to Roe (2002){\cite{Roe2002}}, the displacement of light due to refraction can be determined by 
\begin{equation} \label{eq:1}
R = \frac{n^2-1}{2n^2}\cdot \tan{z_t} \cdot 206265 \cdot 1000 ,
\end{equation}
where $R$ is the refraction in mas, $n$ is the refractive index of air, and $z_t$ is the true zenith distance {\cite{Roe2002}}. Incident DAR can be determined by subtracting the refraction for the two end wavelengths of light. Light disperses along the elevation axis, which can be calculated using the header keywords described in Sec. \ref{subsec:headers}. With direction and magnitude, we have a complete atmospheric dispersion model vector. We then subtracted the model vector from the residual vector to get an inferred correction vector for each spectral datacube that represents the correcting effect of the GPI ADC. Since the Mathar approximation used in $H$ band is invalid in $J$ band, we created a new Mathar approximation valid in $J$ band to model $J$ band incident DAR. The approximation is in Appx. \ref{appx:approx}. Creating $J$ band inferred correction vectors followed the same process as for $H$ band.

\begin{figure}
\begin{center}
\begin{tabular}{c}
\includegraphics[height=8cm]{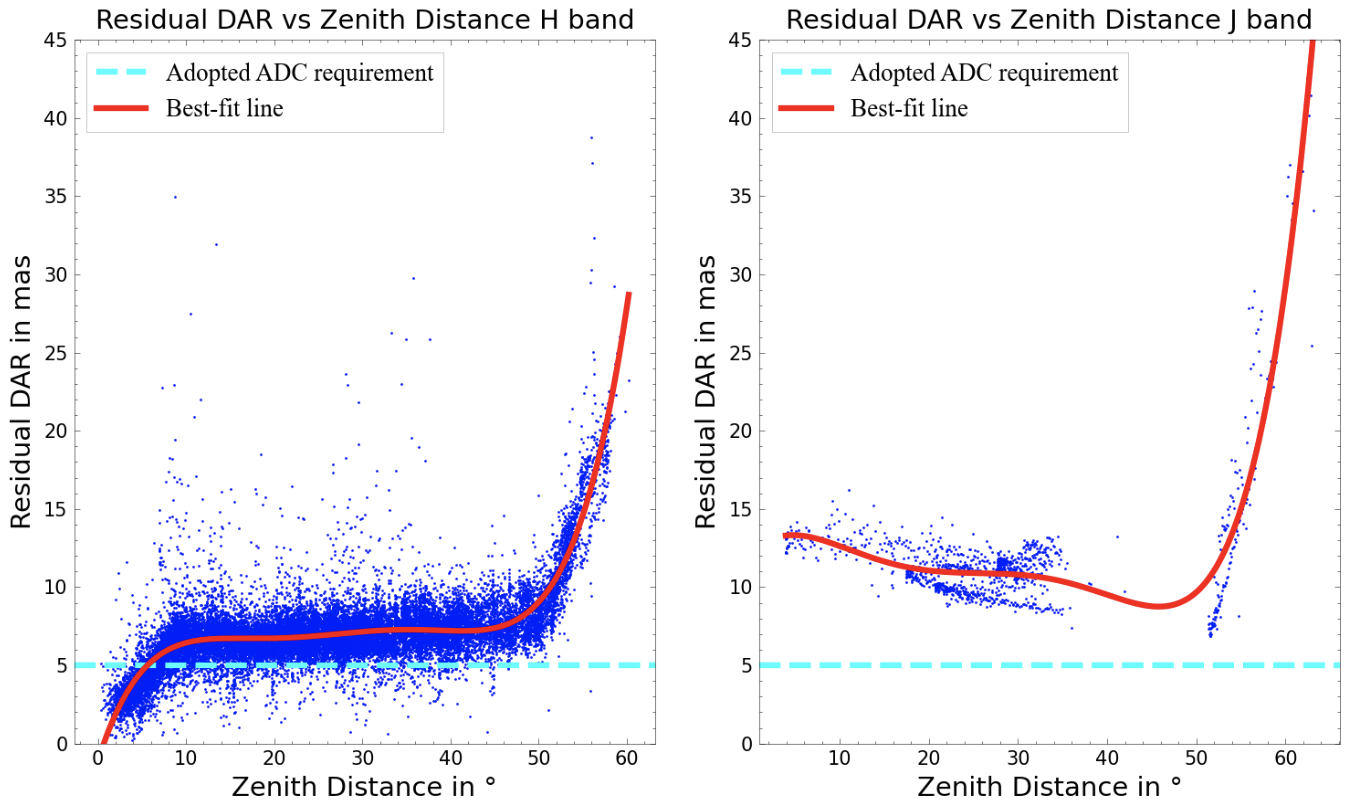}  
\\
(a) \hspace{6cm} (b)
\end{tabular}
\end{center}
\caption 
{ \label{fig:totaldrift} Measured values of the residual DAR-zenith distance relation in (a) $H$ band and (b) $J$ band. Each point is the residual DAR observed in every IFS spectral datacube in the respective bands. Outliers and bad data were pruned as described in Sec. \ref{subsec:prune}. The red lines show the best-fit lines used for the regression described in Sec. \ref{subsec:zenithdistance}. The cyan line shows the GPI ADC's adopted requirement for residual dispersion{\cite{GPIADC}}.} 
\end{figure} 

\section{Analysis of Science Performance}

\subsection{Trends with Zenith Distance}
\label{subsec:zenithdistance}
Zenith distance is the largest determiner of dispersion. Shown in Eq. (\ref{eq:1}), incident DAR is proportional to the tangent of the zenith distance. Residual DAR has a different relationship due to the correction applied by the ADC. The relation between zenith distance and residual DAR in $H$ band can be seen in Fig. \ref{fig:totaldrift}(a). Using the internal calibration source, we found the instrumental DAR to have a median dispersion of 1.29 mas in $H$ band with no appreciable correlation with zenith distance. Thus, it's likely most of the residual DAR is not instrumental. However, since the ADC is located upstream of the internal calibration source, we cannot discount the possibility of the ADC inducing some DAR. At zenith distances beyond 50$^\circ$, the ADC stops making additional corrections and residual DAR increases drastically as the ADC is not designed to work at very high zenith distances.{\cite{GPIADC}} At smaller zenith distances, residual DAR becomes minimal and noise becomes a significant factor. At medium zenith distances, there is consistently about 7 mas in residual DAR, above the 5 mas adopted requirement for the GPI ADC. To view subtler trends in the residual DAR, we fit a fifth-power polynomial to the residual DAR vs zenith distance graph to perform a regression and eliminate the trend, allowing us to find other meaningful relationships with variables such as temperature or humidity which will be discussed in Sec. \ref{subsec:humidity}.

$J$ band data have a different residual DAR-zenith distance trend. Shown in Fig. \ref{fig:totaldrift}(b), $J$ band data have a similar spike in residual DAR at large zenith distances. However, only two objects have been observed at high zenith distances in $J$ band, HR 8799 and V343 Nor. The residual DAR at medium zenith distances is consistently higher in $J$ band than in $H$ band. At small zenith distances, the residual DAR increases slightly instead of approaching zero as it does in $H$ band data. The $J$ band residual dispersion versus zenith distance relationship at low zenith distance is unexpected. Since atmospheric refraction goes to zero at small zenith distances, the GPI ADC must be introducing dispersion at small zenith distances in $J$ band. We will discuss this behavior more in Sec. \ref{subsec:analyzecorrection}. To look at other trends within the data in $J$ band, we similarly performed a regression on this dominant trend. 

$Y$ and $K$ band data show no relation between residual DAR and zenith distance. There is too much noise and too few spectral datacubes for any relation to be revealed. These bands were consequently not used for analysis.

\subsection{Analyzing Parallel and Perpendicular Dispersion }
\label{subsec:analyzecorrection}

\begin{figure}
\begin{center}
\begin{tabular}{c}
\includegraphics[height=19cm]{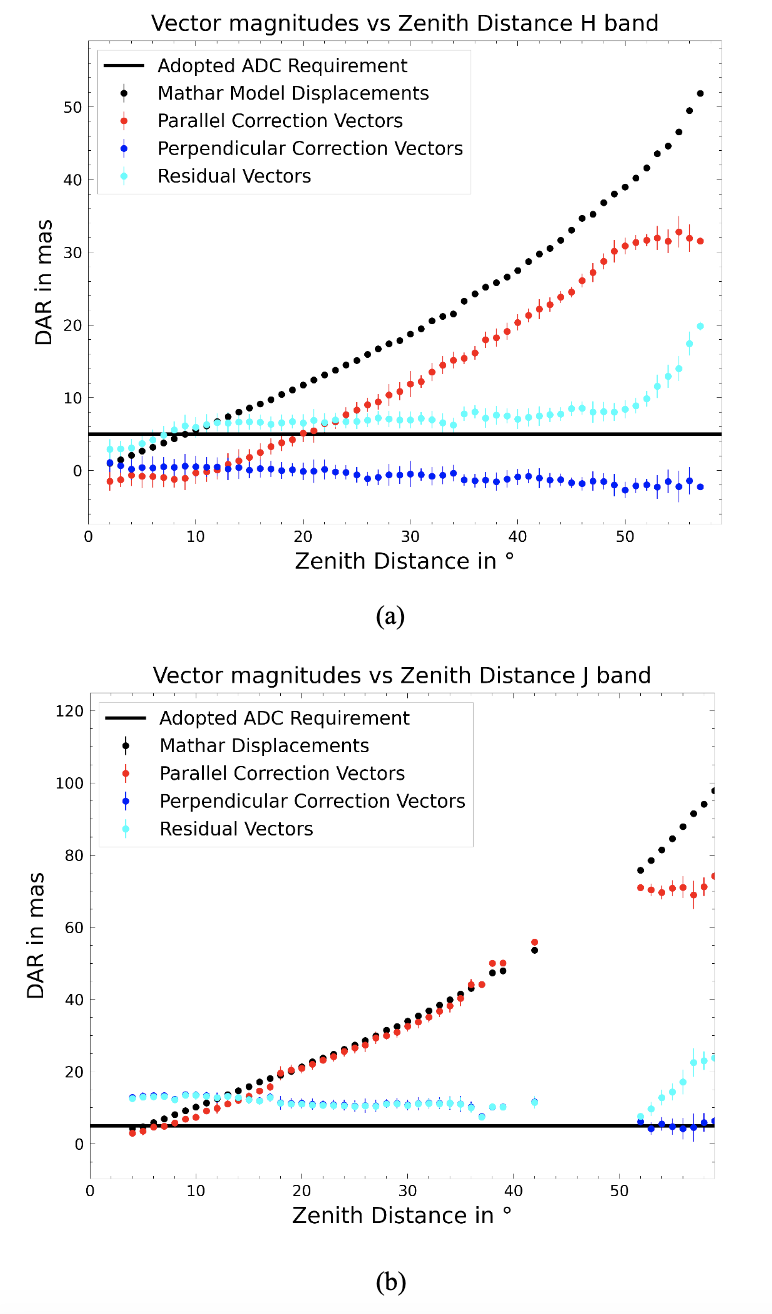}  
\end{tabular}
\end{center}

\caption{ \label{fig:totalperpandpara} Vector magnitudes with the correction vectors broken down into parallel and perpendicular to the zenith in (a) $H$ band and (b) $J$ band. Each data point shows the average of the spectral cubes at each zenith distance. The error bars show the standard deviation. Negative values mean away from the zenith, and thus in the direction of atmospheric DAR, for parallel correction components and 90$^\circ$ clockwise from the zenith for perpendicular components. Incident DAR and residual DAR are shown as positive here for plotting purposes. The perpendicular correction vector component and the residual vectors overlap in $J$ band at zenith distances less than 40$^\circ$.}

\end{figure} 
We broke the residual and correction vectors down into components parallel and perpendicular to the zenith. Parallel correction vectors going away from the zenith, and thus in the direction of atmospheric DAR, are considered negative and perpendicular vectors 90$^\circ$ clockwise from the zenith are considered negative. Even though incident DAR's and residual DAR's direction is opposite the zenith, it is considered positive for plotting purposes in Fig. \ref{fig:totalperpandpara}. Breaking down $H$ band correction vectors into parallel and perpendicular to the zenith components reveals new trends. 

We see in Fig. \ref{fig:totalperpandpara}(a) that at zenith distances 20-50$^\circ$, there is a constant difference between the incident DAR and correction vectors of about 7 mas in $H$ band, above the 5 mas design requirement for the ADC mentioned in Sec. \ref{subsec:GPI}. The ADC is consistently under correcting by a similar magnitude in all spectral datacubes, which is impacting the performance of the GPI. Ideally, there would be no constant difference and the residual DAR would be near zero. The GPI ADC could apply a constant factor to the correction magnitude to account for the consistent undercorrection. Additionally, at zenith distances less than 12$^\circ$, the parallel correction magnitudes plateau in $H$ band, a trend not observed in $J$ band. Interestingly, the plateau coincides with the zenith distances where the incident DAR is less than the 5 mas ADC design requirement, which may explain the plateau's absence in $J$ band. This behavior may be a result of an issue with the GPI ADC control software, but we did not investigate that software in this work. The correction magnitudes' zenith distances appear offset. The correction magnitudes begin going up at 12$^\circ$ zenith distances and have a slightly decreasing zenith distance offset until around 50$^\circ$ zenith distance. The GPI ADC should be correcting at these small zenith distances in $H$ band like it does in $J$ band. 

\begin{figure}
\begin{center}
\begin{tabular}{c}
\includegraphics[height=19cm]{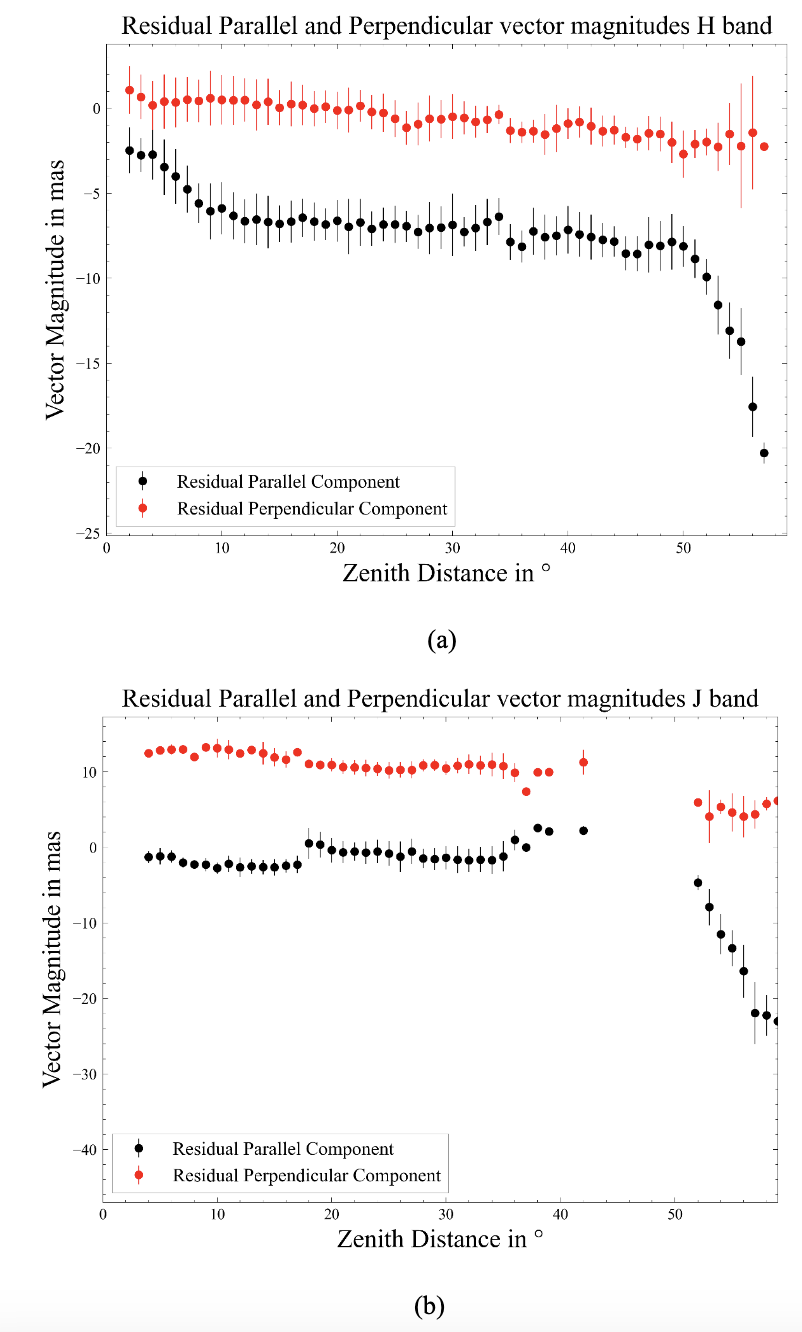}  
\end{tabular}
\end{center}
\caption 
{ \label{fig:residualperpandpara}Residual DAR in (a) $H$ band and (b) $J$ band broken down into parallel and perpendicular to the zenith components. Each data point shows the average of all spectral cubes at that zenith distance. The error bars show the standard deviation. Negative values mean away from the zenith for parallel components and 90$^\circ$ clockwise from the zenith for perpendicular components.} 
\end{figure}

There is also a noticeable trend in the perpendicular dispersion in $H$ band. Since the atmosphere can not induce perpendicular dispersion, the GPI ADC must be introducing the perpendicular dispersion. The perpendicular components gently increase closer to the zenith. From 55$^\circ$ zenith distance to the zenith, the perpendicular component moves in the positive direction 3 mas. Since the perpendicular component starts just short of -3 mas, it has little effect at small zenith distances as it is just above 0. Because of this, it has a small impact on the residual DAR in $H$ band. Figure \ref{fig:residualperpandpara}(a) shows that most of the residual dispersion is in the parallel direction, meaning the perpendicular component only has a small impact while the undercorrection causes the bulk of the issue. However, the perpendicular component's effect in the $J$ band is more profound.

As seen in Fig. \ref{fig:totalperpandpara}(b), the $J$ band parallel correction component is almost perfectly in sync with the predicted dispersion based on the model. There is a small offset that varies greatly but averages about 1.5 mas at zenith distances less than 35$^\circ$ and 2.2 mas at zenith distances less than 15$^\circ$. This offset is minor in comparison to the perpendicular component in $J$ band as seen in \ref{fig:residualperpandpara}(b). Around 86\% of the $\sim11$ mas residual dispersion is in the perpendicular direction. The parallel residual DAR in $J$ band shows the ADC's potential precision. The $J$ band perpendicular correction component has a similar changing trend as $H$ band correction vectors. However, at 55$^\circ$ zenith distances, its perpendicular component is already at positive 5.5 mas. Moving towards the zenith, it increases an additional 7 mas. This results in a large perpendicular component becoming present in the residual DAR. The ratio of the perpendicular magnitude change in $J$ band data to the perpendicular magnitude change in $H$ band data is almost exactly equal to the ratio of the mean $J$ band parallel correction magnitude to the mean $H$ band parallel correction magnitude. We hypothesize this perpendicular dispersion observed in $H$ and $J$ band may be caused by an error in the relative alignment of the GPI ADC prisms, although we defer detailed optical modeling to reproduce this effect for future work. A careful realignment of the two ADC prisms may be necessary for GPI 2 {\cite{Chilcote2020}}.  

\subsection{Trends with Atmospheric Conditions}
\label{subsec:humidity}
\begin{figure}
\begin{center}
\begin{tabular}{c}
\includegraphics[height=18cm]{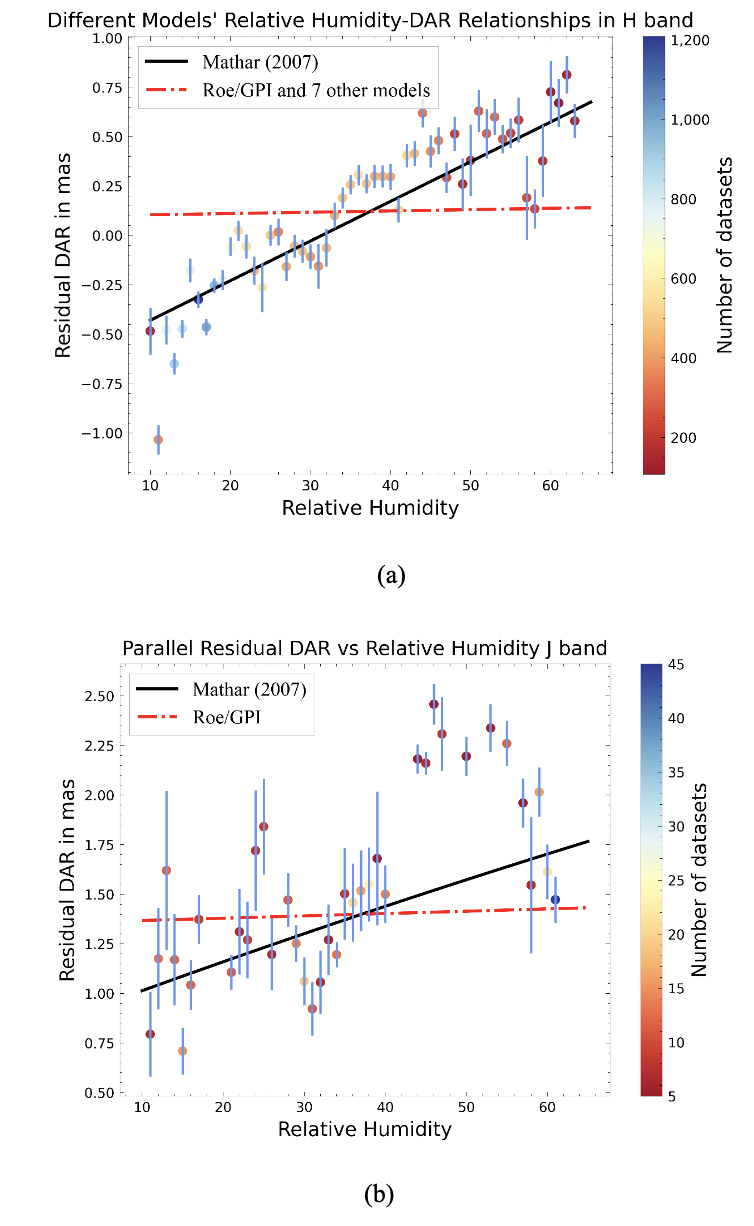}  
\end{tabular}
\end{center}
\caption 
{ \label{fig:humidity}The parallel residual DAR-humidity relationship in (a) $H$ band and (b) $J$ band. The residual DAR here has undergone a regression to remove the residual DAR-zenith distance trend as described in Sec. \ref{subsec:zenithdistance}. The GPI measures humidity at discrete values and each data point shows the average residual DAR for each relative humidity. The error bars show the standard deviation for the average residual dispersion for each relative humidity. The other seven models are all approximately the same as the Roe model. Relative humidities with less than 100 spectral datacubes were thrown out in $H$ band. Relative humidities with less than 5 spectral datacubes were thrown out in $J$ band. We applied a vertical offset to align the models with the residual points so the graph only shows the change of the model as a function of humidity.}  

\end{figure}

\begin{figure}
\begin{center}
\begin{tabular}{c}
\includegraphics[height=18.9cm]{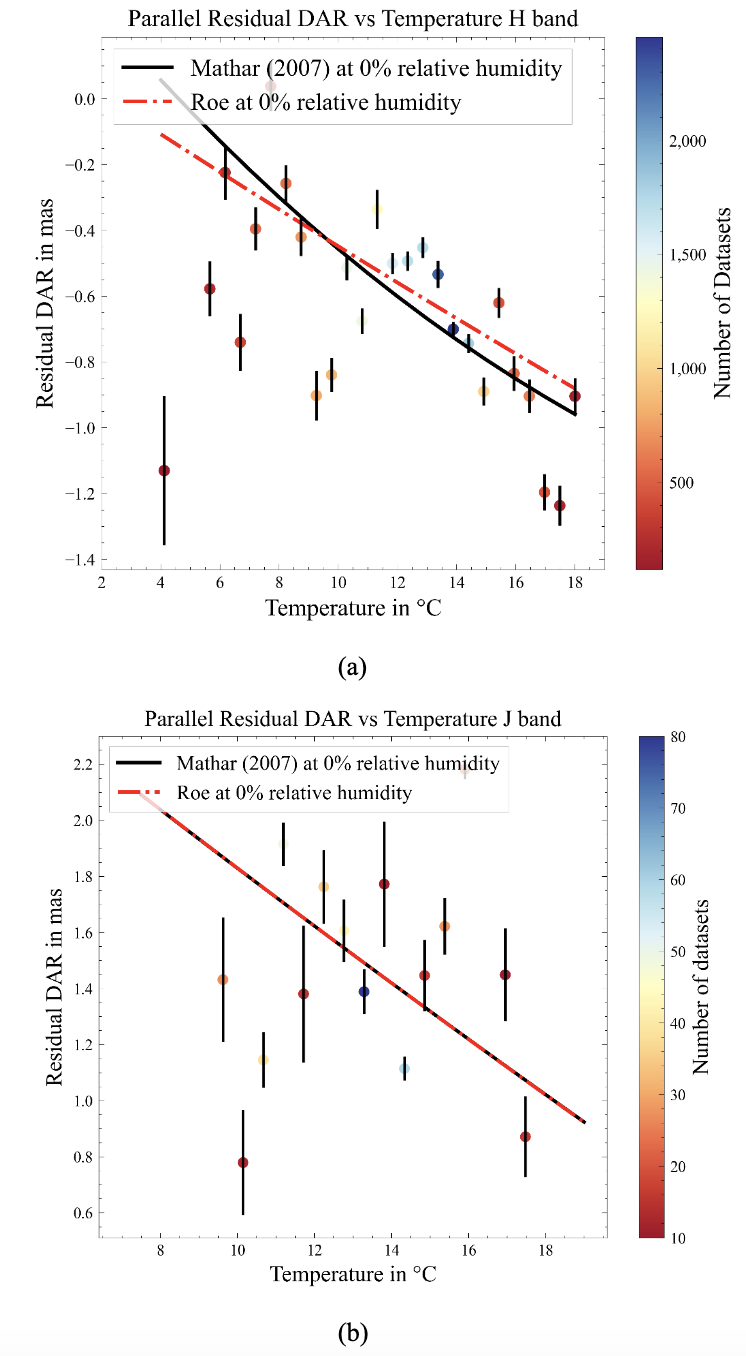}  
\end{tabular}
\end{center}
\caption 
{ \label{fig:temperature}The parallel residual DAR-temperature relationship in (a) $H$ band and (b) $J$ band. The residual DAR here has undergone a regression to remove the residual DAR-zenith distance trend as described in Sec. \ref{subsec:zenithdistance}. The DAR-humidity trend was also removed in both bands. The data were binned such that each point is the average of all spectral datacubes with temperatures in a 0.5$^\circ$C range. The error bars show standard deviation for each bin. Temperature ranges with less than 100 spectral datacubes were thrown out in $H$ band. Temperature ranges with less than 10 spectral datacubes were thrown out in $J$ band. The Mathar and Roe model completely overlap in $J$ band. We applied a vertical offset to align the models with the residual points so the graph only shows the change of the model as a function of humidity. }

\end{figure}

\begin{figure}
\begin{center}
\begin{tabular}{c}
\includegraphics[height=18.9cm]{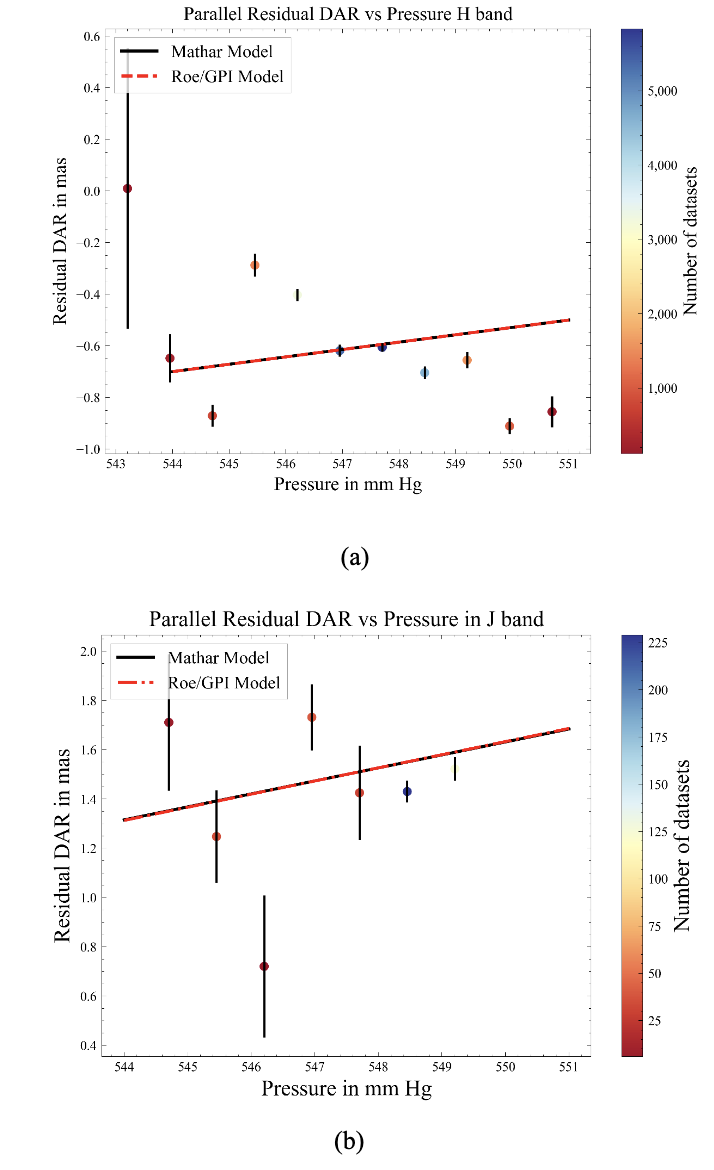}  
\end{tabular}
\end{center}
\caption 
{ \label{fig:pressure}The parallel residual DAR-pressure relationship in (a) $H$ band and (b) $J$ band. The residual DAR here has undergone a regression to remove the residual DAR-zenith distance trend as described in Sec. \ref{subsec:zenithdistance}. The DAR-humidity trend was also removed in both bands. GPI measures pressure at discrete, 0.75 mm Hg intervals. Each data point shows the average residual DAR for the given pressure. The error bars show the standard deviation for each bin. Pressures with less than 100 spectral datacubes were thrown out in $H$ band. Pressures with less than 10 spectral datacubes were thrown out in $J$ band. The Mathar and Roe model completely overlap in both bands. We applied a vertical offset to align the models with the residual points so the graph only shows the change of the model as a function of humidity.} 

\end{figure}
The index of refraction of air, and consequently dispersion, depends on atmospheric conditions including pressure, temperature, and humidity{\cite{Roe2002, Mathar}}. When determining the correction needed, the GPI ADC considers these variables. The GPI ADC uses a model similar to Roe to determine the refractive index of air. However, the Roe model and the GPI ADC do not adequately correct for humidity’s impact on incident DAR. Disagreement between refractive index models regarding humidity's impact on incident DAR has been observed in the past{\cite{roebad}}. Using data after removing the residual DAR-zenith distance trend and breaking it down into parallel and perpendicular to the zenith components, Fig. \ref{fig:humidity}(a) shows the measured parallel residual DAR-humidity relationship and the predicted trend from many different atmospheric models{\cite{HohenkerkAndSinclair, potulski1998, owens1967, edlen1966, edlen1953, ciddor1996, barrellandsears, astroatmo, Mathar, Roe2002}} in $H$ band. The GPI ADC falls short when correcting for humidity. Ideally, there should be no relation between residual DAR and humidity as humidity should be corrected for. However, the relationship has a Pearson correlation coefficient of 0.90. We applied an offset to all the models to align them with the residual points. The slopes show how much each model takes humidity into account. Almost all models, including the Roe and GPI models, greatly underestimate humidity’s impact on DAR. The Mathar model, however, almost perfectly predicts the DAR trend in $H$ band. We confirm with real data that the Mathar model correctly predicts humidity's impact on the refractive index of air in $H$ band while most other refractive index of air models greatly underestimate humidity's impact. The GPI ADC’s model makes negligible corrections for humidity. Using a model that fails to account for humidity causes a decrease in the GPI ADC’s performance. The GPI ADC should use a model for the index of refraction of air that considers humidity to eliminate more dispersion. Subtracting out the observed humidity trend from the residual DAR data, we found that accounting for humidity could reduce dispersion in $H$ band by about 0.54 mas on average, an almost $\sim8\%$ decrease in dispersion. 

$J$ band data corroborate $H$ band data. Figure \ref{fig:humidity}(b) shows the DAR-humidity relation post regression in $J$ band. To avoid other effects appearing in graphs with atmospheric variables, we removed a group of spectral datacubes in the 20 to 40$^\circ$ zenith distance range that seemed to diverge from the residual DAR-zenith distance trend. We're not sure what causes this divergence. This group can be seen in Fig. \ref{fig:totaldrift}(b). The DAR-humidity relation in $J$ band has a Pearson correlation coefficient of 0.68. The Mathar and the Roe model disagree slightly in $J$ band. The data seems to support the Mathar model, but due to the small sample size, it is not entirely clear. The color bar shows some of the points have very few spectral datacubes in $J$ band. The point with the most spectral datacubes in the $J$ band graph has less than half the spectral datacubes than the point with the fewest spectral datacubes in the $H$ band graph, which probably explains why the Pearson correlation coefficient for the residual DAR-humidity relation is lower in $J$ band. Humidity's impact on DAR is minor and the constant undercorrection in $H$ band and the induced perpendicular dispersion in $J$ band have a more significant impact on residual DAR. However, the Roe model's inability to account for humidity when calculating incident DAR reveals a flaw that is impacting the performance of the GPI and possibly other ground-based infrared telescopes. The GPI ADC could switch to using Mathar approximations to properly correct for humidity's impact on DAR. 

Residual DAR does not have as significant of a trend with other atmospheric variables such as temperature or pressure. Figure \ref{fig:temperature} shows the parallel component of residual DAR as a function of temperature in the $J$ and $H$ bands. The data shown underwent a regression to remove trends with both humidity and zenith distance. It has a Pearson correlation coefficient of -0.46 and -0.08 in $H$ band and $J$ band respectively. There is no strong correlation with temperature in either of the bands. The GPI ADC is adequately correcting for temperature. Mathar and Roe both predict similar amounts of DAR with changing temperatures. 

Since the GPI barometer measures pressure in discrete, 0.75 mm Hg ranges and the range of pressures and predicted DAR change is small, it is difficult to analyze trends with pressure. Figure \ref{fig:pressure} shows residual DAR plotted against pressure. The $J$ band data has a Pearson correlation coefficient of 0.11 and the $H$ band has a Pearson correlation coefficient of -0.63. The Roe and the Mathar model completely agree on how DAR changes as a function of pressure. The variance in residual DAR in the pressure graphs is unlikely to be caused by pressure. Thus, there is not a strong trend between residual DAR and pressure.

\section{Conclusion}
The GPI ADC is responsible for correcting the dispersion caused by incident DAR. Its design requirement is less than 5 mas of residual dispersion. We used header data and satellite spots to measure the residual DAR in each spectral datacube and atmospheric models to infer the correction made by the GPI ADC. We analyzed the correction vectors of the GPI ADC to explore potential causes of dispersion. Residual DAR has a strong relation with zenith distance, so we performed a regression to remove that trend to reveal other relationships with residual DAR. This regression allowed us to investigate how different atmospheric variables impact DAR. We found that:

\begin{itemize}
\item the GPI ADC falls short of its design requirements, potentially impacting the coronagraph's performance. 
\item the GPI ADC, the Roe model, and almost all models for the refractive index of air incorrectly predict humidity’s impact on DAR whereas the Mathar model correctly predicts humidity's relationship with DAR. The GPI team should consider using the Mathar approximation to determine the expected DAR.
\item in $H$ band, the correction magnitude has a constant offset from the incident DAR, which causes most of the $H$ band residual dispersion.
\item a perpendicular component in the correction vector is observed in both bands. This perpendicular component is significantly impacting the residual DAR in $J$ band. It may be a result of an issue in the relative alignment of the two ADC prisms.
\end{itemize}

Fixing these issues would improve GPI performance and exoplanet detection. A constant correction can be applied in $H$ band to remedy its consistent undercorrection and the use of the Mathar model could reduce residual DAR caused by humidity. $J$ band DAR is significantly impacted by a perpendicular component. We are not certain about the cause of the perpendicular component, but finding and fixing it would significantly reduce residual DAR in $J$ band. Ray-tracing simulations could shed some light on the $J$ band perpendicular component as the artificial source is downstream of the ADC. Another potentially impractical solution to this issue could be measuring DAR as data are taken to adjust the ADC prisms and eliminate more dispersion. Since the GPI is currently being upgraded, these issues can be addressed.

\appendix

\begin{figure}
\begin{center}
\begin{tabular}{c}
\includegraphics[height=9cm]{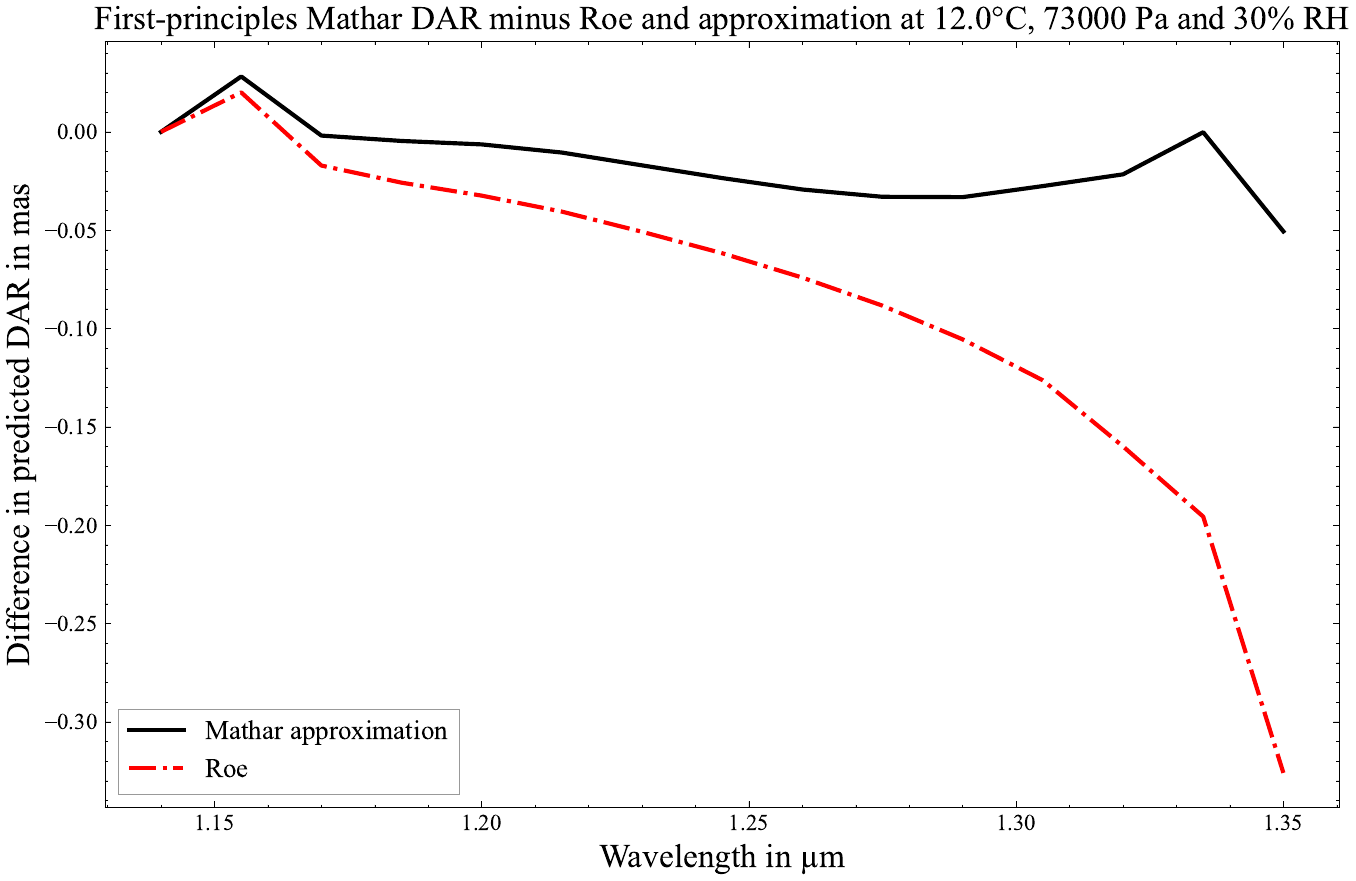}  
\end{tabular}
\end{center}
\caption 
{ \label{fig:approx}Difference in predicted $J$ band DAR of the first-principles Mathar model, Mathar approximation, and Roe model at the average atmospheric conditions at the Gemini South Observatory. The graph shows the Roe model and Mathar approximation subtracted from the first-principles Mathar model, so negative values signal an overcorrection.} 

\end{figure}
\section{Mathar Approximation from 0.7 $\mu$m to 1.36 $\mu$m}
\label{appx:approx}
We created an approximation of the first-principles Mathar model in the 0.7 $\mu$m to 1.36 $\mu$m range. This approximation is valid from -10$^\circ$C to 20$^\circ$C, 50000 Pa to 83250 Pa, and 5\% relative humidity to 95\% relative humidity. It uses reference values of $T_{ref} = 280.65$$^\circ$K, $p_{ref} = 66625$ Pa, $H_{ref} = 50\%$, and $\lambda_{ref}=0.77$$\mu$m. The approximation is in the format described in Mathar (2007){\cite{Mathar}} except we use wavelengths instead of wavenumbers. The approximation in this wavelength range is 
\begin{equation} \label{eq:2}
n-1 = \sum_{i=0,1,2...} c_i(T, p, H)(\lambda-\lambda_{ref})^i,
\end{equation}
where 
\begin{align*} \label{eq:3}
c_i = &c_{iref} \\
&+ c_{iT}(\frac{1}{T}-\frac{1}{T_{ref}}) + c_{iTT} (\frac{1}{T}-\frac{1}{T_{ref}})^2  \\
&+ c_{iH}(H-H_{ref}) + c_{iHH}(H-H_{ref})^2  \\
&+ c_{ip}(p-p_{ref}) + c_{ipp}(p-p_{ref})^2  \\
&+ c_{iTH}(\frac{1}{T}-\frac{1}{T_{ref}})(H-H_{ref})  \\
&+ c_{iTp}(\frac{1}{T}-\frac{1}{T_{ref}})(p-p_{ref}) \\
&+ c_{iHp}(H-H_{ref})(p-p_{ref}).\\
\end{align*}
Table \ref{tab:thetable} shows the fitting coefficients and their units for the Taylor expansion.\footnote{The new Mathar approximation integrated with two other approximations is available on GitHub at \href{https://github.com/semaphoreP/nair/blob/master/nair.py}{https://github.com/semaphoreP/nair/blob/master/nair.py}   }

Across the atmospheric variables where the approximation is valid, the Roe model's predicted $J$ band DAR is, on average, 0.42 mas off from the first-principles Mathar model while our new Mathar approximation is only off by an average of 0.13 mas. At the average conditions for the Gemini South Observatory (12$^\circ$C, 30\% relative humidity, and 73,000 Pa), the Roe model predicts 0.33 mas more incident DAR than the first-principles Mathar model while the approximation predicts only 0.05 mas more incident DAR, as seen in Fig. \ref{fig:approx}. The new Mathar approximation better models humidity's effect on DAR in $J$ band and also enables near-infrared instruments to use Mathar approximations for all of their bands.

\begin{table}
\caption{Coefficients for the Mathar approximation valid from 0.7 $\mu$m to 1.36 $\mu$m using reference values of $T_{ref} = 280.65$$^\circ$K, $p_{ref} = 66625$ Pa, $H_{ref} = 50\%$, and $\lambda_{ref}=0.77$$\mu$m.}
\label{tab:thetable}
\fontsize{9pt}{20pt}
\begin{center}

\begin{tabular}{||c c c c c c||}

 \hline
 $i$ & $c_{iref}$ / $[1/{\mu m}^{i}]$ & $c_{iT}$ / $[K/{\mu m}^i]$ & $c_{iTT}$ / $[K^2/{\mu m}^i]$ & $c_{iH}$ / $[1/({\mu m}^i\%)]$& $c_{iHH}$ / $[1/({\mu m}^i\%^2)]$\\  

 \hline\hline
 0 & $1.85566259\times10^{-4}$ & $5.33344343\times10^{-2}$ & $4.37191645\times10^{-6}$ & $-5.29847992\times10^{-9}$ & $ 1.72638330\times10^{-13}$\\ 
 \hline
 1 & $-4.68511206\times10^{-6}$ & $-1.24712782\times10^{-3}$ & $-6.25121335\times10^{-8}$ & $-3.13820651\times10^{-10}$ & $1.61933914\times10^{-12}$\\
 \hline
 2 & $9.19681919\times10^{-6}$ & $2.33119745\times10^{-3}$ & $1.63938942\times10^{-7}$ & $4.69827651\times10^{-10}$ & $-5.64003179\times10^{-12}$\\
 \hline
 3 & $-1.44638085\times10^{-5}$ & $-2.32913516\times10^{-3}$ & $-2.11103761\times10^{-7}$ & $-3.50677283\times10^{-9}$ & $-2.62670875\times10^{-12}$\\
 \hline
 4 & $1.52286899\times10^{-5}$ & $1.75945139\times10^{-6}$ & $-1.52898469\times10^{-8}$ & $9.63769669\times10^{-9}$&$1.21144700\times10^{-11}$\\
 \hline
 5 & $-7.42131053\times10^{-6}$ & $1.51989359\times10^{-3}$ & $1.13124404\times10^{-7}$ & $-9.13487764\times10^{-9}$ & $4.26582641\times10^{-12}$\\ 
 \hline
\end{tabular}
\end{center}

\fontsize{9pt}{20pt}
\begin{center}
\begin{tabular}{||c c c c c c||} 
 \hline
 $i$ & $c_{ip}$ / $[1/({\mu m}^iPa)]$ & $c_{ipp}$ / $[1/({\mu m}^iPa^2)]$ & $c_{iTH}$ / $[K/({\mu m}^i\%)]$& $c_{iTp}$  / $[K/({\mu m}^iPa)]$ & $c_{iHp}$  / $[1/({\mu m}^iPa\%)]$\\  
 \hline\hline
 0 & $2.78974970\times10^{-9}$ & $2.26729683\times10^{-17}$ & $2.12082170\times10^{-5}$ & $7.85881100\times10^{-7}$ & $ -1.40967131\times10^{-16}$\\ 
 \hline
 1 & $-7.00536198\times10^{-11}$ & $7.56136386\times10^{-18}$ & $1.29405965\times10^{-6}$ & $-1.97232615\times10^{-8}$ & $1.64663205\times10^{-18}$ \\
 \hline
 2 & $1.37565581\times10^{-10}$ & $-4.20128342\times10^{-17}$ & $-6.13606755\times10^{-6}$ & $3.87305157\times10^{-8}$ & $-7.48099499\times10^{-18}$\\
 \hline
 3 & $-2.14757969\times10^{-10}$ & $2.08166817\times10^{-17}$ & $4.29222261\times10^{-5}$ & $-6.04645236\times10^{-8}$ & $8.67361738\times10^{-18}$ \\
 \hline
 4 & $2.22197137\times10^{-10}$ & $2.94902991\times10^{-17}$ & $-1.04934521\times10^{-4}$& $6.25595229\times10^{-8}$ & $-6.93889390\times10^{-18}$\\
 \hline
 5 & $-1.04766954\times10^{-10}$ & $6.24500451\times10^{-17}$ & $8.65209674\times10^{-5}$& $-2.94970993\times10^{-8}$ & $-1.73472348\times10^{-18}$\\ 
 \hline 
\end{tabular}
\end{center}
\end{table}
\fontsize{12pt}{20pt}

\bigskip
\noindent
\textit{Code, Data, and Materials Availability}

\medskip
\noindent
The data used in this analysis will be publicly released as part of the final GPI Exoplanet Survey data release, which is currently under progress at the time of writing of this manuscript. In the meantime, the specific data and code can be requested from the corresponding author.

\acknowledgments
Based on observations obtained at the international Gemini Observatory, a program of NSF’s NOIRLab, which is managed by the Association of Universities for Research in Astronomy (AURA) under a cooperative agreement with the National Science Foundation on behalf of the Gemini Observatory partnership: the National Science Foundation (United States), National Research Council (Canada), Agencia Nacional de Investigaci\'{o}n y Desarrollo (Chile), Ministerio de Ciencia, Tecnolog\'{i}a e Innovaci\'{o}n (Argentina), Minist\'{e}rio da Ci\^{e}ncia, Tecnologia, Inova\c{c}\~{o}es e Comunica\c{c}\~{o}es (Brazil), and Korea Astronomy and Space Science Institute (Republic of Korea). We'd also like to thank Jeffrey Chilcote for discussions of the manuscript.
This research made use of Astropy, a community-developed core Python package for Astronomy {\cite{Astropy2013,Astropy2018}}.  
Part of the research was carried out at the Jet Propulsion Laboratory, California Institute of Technology, under a contract with the National Aeronautics and Space Administration (80NM0018D0004).

\nolinenumbers
\bibliography{bib}   
Author biographies are unavailable. 
\bibliographystyle{spiejour}

\listoffigures
\listoftables

\end{spacing}
\end{document}